\documentclass{ptephy_v1}
\preprintnumber{XXXX-XXXX} 

\graphicspath{{./picture/}}
\usepackage{here}

\begin{document}

\title{Evaluation of  {loop formation dynamics} in a chromatin fiber  {during chromosome condensation}}


\author{Hiroshi Yokota}
\author[1, 2]{Masashi Tachikawa}

\affil{Interdisciplinary Theoretical and Mathematical Science Program (iTHEMS), RIKEN, Saitama 351-0198, Japan \email{hiroshi.yokota@riken.jp}}
\affil[2]{Institute for Frontier Life and Medical Sciences, Kyoto University, Shogoin Kawahara-cho, Sakyo-ku, Kyoto 606-8507 Japan}




\begin{abstract}%
Chromatin fibers composed of DNA and proteins fold into consecutive loops to form rod-shaped chromosome in mitosis. 
Although  {the loop growth dynamics is investigated in several studies},  {its detailed processes are unclear.} 
Here, we  {describe the time evolution of the loop length for thermal-driven loop growth processes as an iterative map by calculating physical quantities involved in the processes.} 
We  {quantify} the energy during  {the} chromatin loop formation by calculating the free energies of unlooped and looped chromatins  {using the Domb-Joyce model of a lattice polymer chain incorporating the bending elasticity for thermal-driven loop growth processes.} 
The excluded volume interaction among loops is integrated by employing the mean-field theory.
We compare the loop formation energy with the thermal energy and evaluate the growth of loop length via  {thermal fluctuation}.
By assuming the dependence of the excluded volume parameter on the loop length,  {we construct an iterative map for the loop growth dynamics.}
The map  {demonstrates} that the growth length of the loop for a single reaction cycle decreases with time to reach the condensin size, where the loop growth dynamics can be less stochastic and be regarded as direct power stroke of condensin as a kind of motor proteins.
\end{abstract}

\subjectindex{xxxx, xxx}

\maketitle

\section{Introduction}
\begin{figure}[tbp]
  \begin{center}
    \includegraphics[width=13cm]{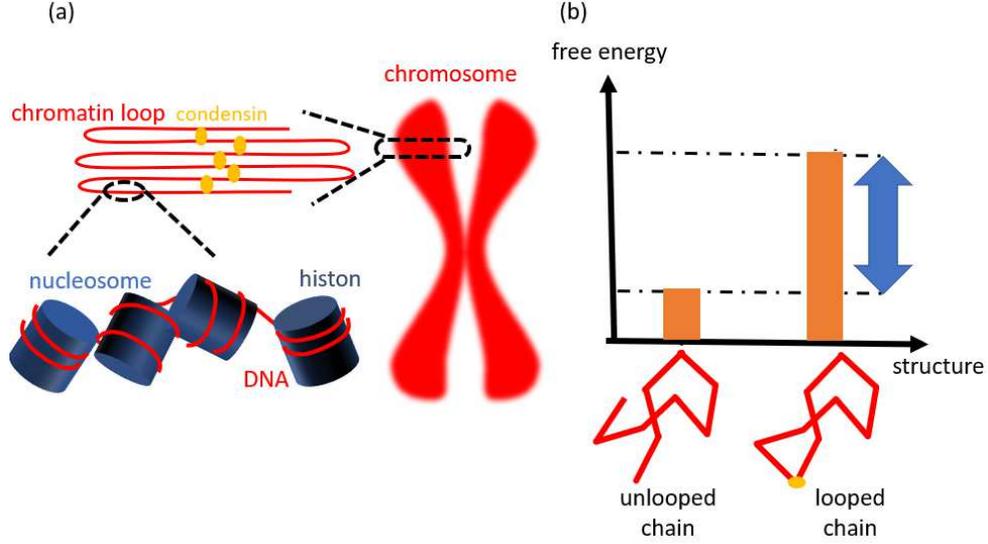}
    \caption{
    (a) Schematic of a chromosome composed of consecutive chromatin loops
    (b)  {One of the physical quantities including the thermal-driving scenario.}
     {The energy source for loop formation can be evaluated by using the free energy difference before and after chromatin loop (s) formation.}
    }
    \label{fig:motivation}
 \end{center}
\end{figure}

\hspace{6mm} 
The genomitc DNA in eukaryotic cells associates with histone proteins to form more flexible and compact fiber called chromatin.
Before mitosis, the chromatin fibers in eukaryotic cells fold further into consecutive loop structures and condense into rod-like chromosomes (see Fig.~\ref{fig:motivation} (a)).
The stiffness of  {the} chromosomes results from the excluded volume interaction between chromatin loops\cite{Goloborodko}.
One of the essential molecules for  {the} chromatin loop formation is a five-subunit protein complex named condensin, which belongs to the highly conserved family of SMC complexes.

 {Although} the mechanism and detailed dynamics of loop formation have been extensively studied, they remain incompletely understood.
One of the most promising hypotheses is that of loop extrusion proposed by Alipour et al.\cite{loop_extruding_static}.
In this hypothesis, condensin binds to two neighboring sites in the chromatin fiber and extrudes (pushes) to form and enlarge the chromatin loop. 
Loop extrusion has been theoretically modeled based on stochastic\cite{Goloborodko_lattice} and coarse-grained molecular dynamics simulations\cite{Goloborodko, Sakai2018}.
Ganji {\it et al.} experimentally observed the loop extrusion activity of condensin on bare DNA\cite{Ganji}.

 {Another important debate topic is the detailed processes for the loop formation, {\it i.e.}, the generation and growth of a chromatin loop by condensin}.
 {From the perspective of the origin of the driving force, t}wo possible candidates are considered. 
One is the direct power stroke of condensin as a motor protein coupled with ATP hydrolysis.
Recent experiments revealed the ATP-dependent translocation and loop formation activity of condensins along DNA \cite{Terakawa, Ganji}; Terakawa et al.\cite{Terakawa} proposed some motor activity models of condensins.
 {However, the direct evidence for the condensin power stroke has not been observed.}
 {Despite the studies on condensin activities, the detailed process model involving the direct power stroke has not been proposed so far.} 

 {Then, some researchers have focused on the other candidate that describes the thermal-driven loop formation.}
 {In these processes, the growth of the loop length per unit time is stochastically determined owing to} thermal fluctuation.
 {A} model proposed by Marko {\it et al.}  {(thermal-driving scenario)}\cite{DNA_capture_Marko} outlines how thermal fluctuations can act as the driving force for the motor activity of condensins. 
According to this model, the DNA fiber and the bacterial SMC protein undergo a cyclic reaction involving conformational changes in the SMC protein and several DNA forms captured by the SMC protein.
During this reaction, the SMC protein binds to DNA and captures a loop that is stochastically formed nearby. Then, the SMC protein releases the former DNA-binding site and rebinds to one of the looped DNA sites, as it did initially.
ATP hydrolysis prevents the reverse reaction and allows a unidirectional movement along DNA.
The net rate of this cyclic reaction $k_{\rm cycle}$ depends on physical quantities, including the rate of ATP hydrolysis by condensin.
Although the model proposed by Marko et al.\cite{DNA_capture_Marko} describes the translocation of the bacterial SMC complex along DNA, the basic idea applies to chromatin loop formation by condensins. 

 {Some physical quantities included in the thermal-driving scenario have been calculated in previous studies.}
 {In the thermal-driving scenario, the growth of the loop length per unit time can be obtained from the free energies of looped and unlooped chromatin fibers because its difference is supposed to be fulfilled by the thermal fluctuation energy (Fig.~\ref{fig:motivation} (b)). }
 {Several researchers reported on the free energy of loop conformation for various models\cite{HWLC_J_factor, bead_spring_J_factor, slip_link_cohesin2, slip_link_cohesin1, brackley2016simulated, brackley2018extrusion, DJ}}.
The free energy of circular DNA is frequently calculated using a helical worm-like chain model based on a continuous polymer chain model, including the curvature and  {the} torsion.
 {In contrast}, the bead-spring model with bending elasticity was used to calculate the free energy of the chromatin or DNA loop\cite{slip_link_cohesin2, slip_link_cohesin1, brackley2016simulated, brackley2018extrusion}.
 {While, the excluded volume interaction was not considered in these studies,}  {Domb and Joyce described the statistical properties of the flexible looped and unlooped chains with and without the excluded volume effect by proposing the concept of a lattice polymer model.}
 {Especially, the statistical properties of the flexible chain is comprehensively considered in the Domb-Joyce model\cite{DJ, RG_and_MC_for_DJ, RW_lattice}.}
 {However, no previous study has investigated the total chromosome condensation processes via the loop formation dynamics using the thermal-driving scenario.}

In this study,  {we reproduced the time evolution of the loop length comparable to experiments by constructing}  {an iterative map}  {based on the thermal-driving scenario.}
 {This}  {iterative map} {is composed of the free energy of semi-flexible loop conformation, the excluded volume effect, and the progression of the excluded volume interaction along with the chromosome condensation.
A single step growth of the loop is represented by}  {the}  {iterative map.}
 {The evaluation of  {the loop growth length} indicates that the thermal-driving scenario is sufficient to reproduce the chromosome condensation.}
 {We also evaluated the size of the single step growth in comparison with the size of condensin and provided a condensation for the various scenarios.}
\section{Model and result}
\subsection{Polymer chain model without loop constraint}\label{sec:polymer_chain}
\hspace{6mm} {This section describes the chromatin/DNA fiber by using a face-centered cubic lattice model for a semi-flexible polymer which incorporates the bending elasticity in the lattice model for a flexible polymer proposed by Domb and Joyce\cite{DJ}.}
The microscopic energy, which characterizes the bending stiffness of the chain, is defined in this lattice model.
The bending elasticity specifies the difference between fibers.
Then, the free energy of the polymer chain without a loop constraint is calculated using the transfer matrix method.
\subsubsection{Lattice chain model and its microscopic energy}
\begin{figure}[t]
  \begin{center}
    \includegraphics[width=13.5cm]{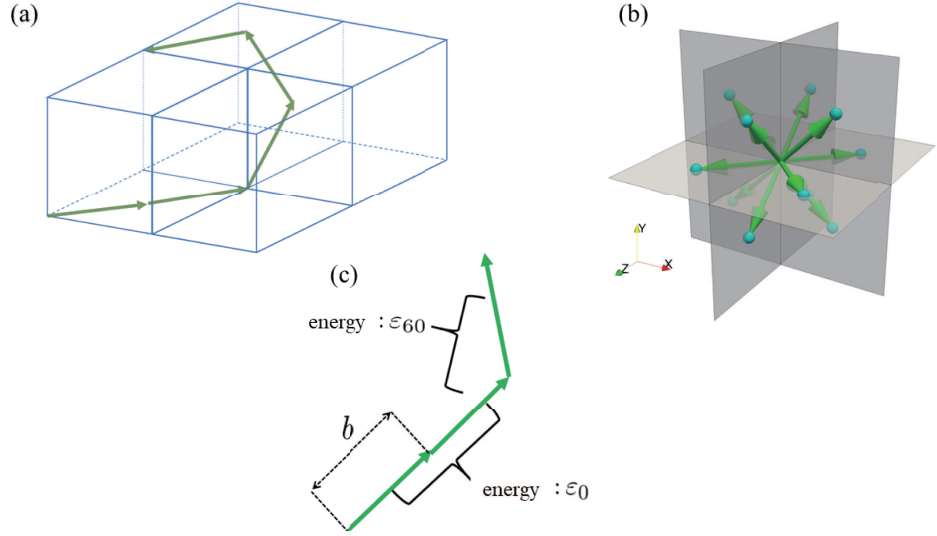}
    \caption{
    (a) The chain model (lattice model).
    The lattice model is depicted as walks on the face-centered cubic lattice, where the end of each segment is at a corner or a face-centered point.
    (b) Unit vectors of the face-centered cubic lattice.
    (c) Bending energy of the chain.
    When the angle between two consecutive segments is $0^\circ$ or $60^\circ$, the chain obtains energy $\varepsilon_0$ and $\varepsilon_{60}$, respectively   {(the Domb-Joyce model with bending elasticity)}. 
    $b$ denotes the size of a segment in the chain.
    }
    \label{fig:model_for_polymer_chain}
 \end{center}
\end{figure}

\hspace{6mm}A chromatin/DNA fiber is described by a {lattice model} composed of consecutive rod-like segments with length $b$ on a face-centered cubic lattice\cite{Yokota_induction}.
The direction of each segment corresponds to one of the 12 unit vectors of the face-centered cubic lattice ; $\mbox{\boldmath$e$}^{(1)}=(b, b, 0)/\sqrt{2}, \mbox{\boldmath$e$}^{(2)}=(-b, b, 0)/\sqrt{2}, \mbox{\boldmath$e$}^{(3)}=(b, 0, b)/\sqrt{2}, \mbox{\boldmath$e$}^{(4)}=(-b, 0, b)/\sqrt{2}, \mbox{\boldmath$e$}^{(5)}=(0, b, b)/\sqrt{2}, \mbox{\boldmath$e$}^{(6)}=(0, -b, b)/\sqrt{2}$ and $\mbox{\boldmath$e$}^{(\eta)}=-\mbox{\boldmath$e$}^{(\eta-6)}$ ($\eta=7, 8, \cdots , 12$) (see Figs.~\ref{fig:model_for_polymer_chain} (a) and (b) ).

Here, we present the microscopic bending energy for the lattice model and calculate the free energy.
In  {this lattice model}, two consecutive segments $0^\circ$ and $60^\circ$ store the microscopic energies $\varepsilon_0$ and $\varepsilon_{60}$, respectively ( {the Domb-Joyce model with bending elasticity}, see Fig.~\ref{fig:model_for_polymer_chain} (c) ) .
Moreover, the storing energy for either of the other angles is $\varepsilon_{\rm other}=\infty$.
This means that these angles are energetically prohibited. 
The statistical weights of the angles $0^\circ$ and $60^\circ$ are $\exp{\left[-\varepsilon_0/(k_{\rm B}T) \right]}$ and $\exp{\left[-\varepsilon_{60}/(k_{\rm B}T) \right]}$, respectively, where $k_{\rm B}$ is the Boltzmann constant and $T$ is the temperature.
The statistical weights of the other angles are $\exp{\left[-\varepsilon_{\rm other}/(k_{\rm B}T) \right]}=\exp{\left[-\infty/(k_{\rm B}T) \right]}=0$.
The energy difference $\Delta \varepsilon \equiv \varepsilon_{60}-\varepsilon_{0}$ quantitatively describes the chain stiffness.
Here, we referred to the relation of the two consecutive segments that form an angle of $0^ \circ$ as a “parallel state” (or parallel conformation), while the $60^\circ$ state is termed as “bending state” (or bending conformation). 
\subsubsection{Transfer matrix method}
\hspace{6mm}To calculate the partition function of  {the Domb-Joyce model with bending elasticity}, we use the transfer matrix method.
We define  
\begin{align}
  \delta =\exp{\left[-\frac{\Delta\varepsilon}{k_{\rm B}T} \right]}
\end{align}
Then, we introduce the transfer matrix\cite{Yokota_induction} as follows
\begin{align}
  \mathcal{T}_{\eta\xi}=\begin{cases}
    1 \ \ \
    ({\rm if\ {parallel\ conformation}})\\
    \delta \ \ \
    ({\rm if\ {bending\ conformation}})\\
    0 \ \ \ ({\rm otherwise}), \\
  \end{cases}
  \label{eqn:transfermatrix}
\end{align}
where $\eta$ and $\xi$ ($\eta, \xi = 1, 2, \cdots , 12$) are the indices of the orientation of two consecutive segments, respectively.
For example, $\mathcal{T}_{11} = 1$ represents the statistical weight of the parallel state, where both the $n$-th and $n+1$-th segment  {point} in the direction parallel to $\mbox{\boldmath$e$}^{(1)}$.
Using eqn.~(\ref{eqn:transfermatrix}), we calculate the partition function of the chain composed of $N+1$ segments of the unlooped state
\begin{align}
  Z_{\rm free}(T, N+1)=\sum_{\eta, \xi=1}^{12} \left( \mathcal{T}^N\right)_{\eta\xi} = 12 (1+4\delta)^N \label{eqn:Z_free}. 
\end{align}
The prefactor $12$ shows that the orientation of the initial segment can be directed to any of the 12 vectors $\mbox{\boldmath$e$}^{(1)}, \mbox{\boldmath$e$}^{(2)} \cdots , \mbox{\boldmath$e$}^{(12)}$.
Factor $1+4\delta$ is derived from the fact that two consecutive segments can take one parallel state or one of the four bending states.  
\subsection{Loop constraint}
\hspace{6mm}Loop formation restricts the conformation of a polymer chain and therefore increases its free energy.
We calculate the partition function of a looped polymer chain with the end-to-end distance equal to zero.
 {Because} the conformational free energy calculated using the transfer matrix only refers to the orientation of the segment, we have to integrate the degrees of freedom of the position of each segment to evaluate the end-to-end distance.
As the end-to-end vector of the chain is obtained using the sum of the $N+1$ segment vectors, we calculate the statistical weight
\begin{align}
  Q(0, \eta, \mbox{\boldmath$r$}^{(0)}; N, \xi, \mbox{\boldmath$r$}^{(N)})&=
  \sum_{\left\{ \mbox{\boldmath$b$}_n \right\}}\left( \mathcal{T}^N \right)_{\eta\xi}
  \delta\left(\mbox{\boldmath$r$}^{(N)}-\mbox{\boldmath$r$}^{(0)}-\left\{ \frac{\mbox{\boldmath$b$}_{0}}{2}+\sum_{i=1}^{N-1}\mbox{\boldmath$b$}_{i} + \frac{\mbox{\boldmath$b$}_{N}}{2}\right\} \right),\label{eqn:definition_of_path_integral}
\end{align}
where the $0$-th segment at position $\mbox{\boldmath$r$}^{(0)}$ and the $N$-th segment at position $\mbox{\boldmath$r$}^{(N)}$ point to $\mbox{\boldmath$e$}^{(\eta)}$ and $\mbox{\boldmath$e$}^{(\xi)}$, respectively.
 {Moreover, $\mbox{\boldmath$b$}_n$ represents the orientation vector of the $n$th segment in a certain conformation of the chain.}
 {$\sum_{\left\{ \mbox{\boldmath$b$}_n \right\}}$ denotes the summation with respect to all conformations of the chain.}
By using the Fourier expression of the $\delta$ function, eqn.~(\ref{eqn:definition_of_path_integral}) can be rewritten as
\begin{align}
  Q(0, \eta, \mbox{\boldmath$r$}; N, \xi, \mbox{\boldmath$r$}^\prime)&=
  \int d\mbox{\boldmath$q$} \mathcal{\tilde{T}}_{\eta\xi}^N(\mbox{\boldmath$q$})
  \exp{\left[i \mbox{\boldmath$q$}\cdot 
  \left(\mbox{\boldmath$r$}^\prime-\mbox{\boldmath$r$}\right) \right]}\label{eqn:path_integral},
\end{align}
where,  
\begin{align}
  \mathcal{\tilde{T}}_{\eta\xi}(\mbox{\boldmath$q$})&=
  \exp{\left[-\frac{i}{2}\mbox{\boldmath$q$}\cdot 
  \mbox{\boldmath$e$}^{(\eta)} \right]}
  \mathcal{T}_{\eta\xi}
  \exp{\left[-\frac{i}{2}\mbox{\boldmath$q$}\cdot 
  \mbox{\boldmath$e$}^{(\xi)} \right]}.
\end{align}
We obtain the loop structure of the chain from the chain conformation under constrained conditions, where the ends of the chain have the same position.
Therefore, the partition function of the loop structure of a chain composed of $N+1$ segments $Z_{\rm loop}(T, N+1)$, is calculated as
\begin{align}
  Z_{\rm loop}(T, N+1)&=\sum_{\eta, \xi=1}^{12}Q(0, \eta, \mbox{\boldmath$r$}; N, \xi, \mbox{\boldmath$r$})
 =\sum_{\eta, \xi=1}^{12}  \int d\mbox{\boldmath$q$} \mathcal{\tilde{T}}_{\eta\xi}^N(\mbox{\boldmath$q$})\label{eqn:Z_loop}.
\end{align}
The free energy of a phantom loop, $F_0$, is calculated as
\begin{align}
  F_{0}=-k_{\rm B}T\ln{\left[\frac{Z_{\rm loop}}{Z_{\rm free}}\right]}\label{eqn:ideal_loop},
\end{align}
where its original point is the value of the unlooped state.

We also calculate the free energy of multiple phantom loops as
\begin{align}
  F_{0}^{(\rm multi)}=\alpha_{\rm all} \times F_{0},
\end{align}
where we assume each loop to be composed of $N+1$ segments, and $\alpha_{\rm all}$ is the number of loops in the system.
\subsection{Excluded volume interaction}\label{sec:mean_field_theory}
\hspace{6mm}Chromosome condensation, along with loop formation, increases the interaction of the excluded volume between loops, which will affect the state of polymers.
We consider the excluded volume interaction among chromatin loops and  {integrate} it into the free energy of the lattice model. 
We adopt the mean-field theory, in which the statistical properties of multiple interacting loops are determined using approximate calculations of the statistical properties of a single loop under a potential field.
To introduce the excluded volume interaction between segments using mean-field theory, we start with a Hamiltonian: 
\begin{align}
  \mathcal{\hat{H}}(\Gamma)&=\mathcal{\hat{H}}_{\rm ideal}(\Gamma) + \sum_{i, j} \mathcal{\hat{W}}(\mbox{\boldmath$r$}^{(i)}, \mbox{\boldmath$r$}^{( {j})}), \label{eqn:hamiltonian}
\end{align}
where the first term describes the Hamiltonian of an ideal system (corresponding to a single polymer without an excluded volume interaction) and the other term refers to the interaction term (corresponding to the interaction between segments).
Moreover $\hat{\cdots}$ indicates that the quantity $\cdots$ is a function of the phase space $\Gamma=\left\{ \mbox{\boldmath$r$}^{(0)}, \mbox{\boldmath$b$}^{(0)}, \mbox{\boldmath$b$}^{(1)}, \ldots,  \mbox{\boldmath$b$}^{(N)} \right\}$.
Equation~(\ref{eqn:hamiltonian}) is rewritten as
\begin{align}
  \mathcal{\hat{H}}&=\mathcal{\hat{H}}_{\rm ideal}+\iint d\mbox{\boldmath$r$}d\mbox{\boldmath$r$}^\prime
  \mathcal{W} (\mbox{\boldmath$r$}, \mbox{\boldmath$r$}^\prime) 
  \hat{\phi}(\mbox{\boldmath$r$})\hat{\phi}(\mbox{\boldmath$r$}^\prime), \label{eqn:hamiltonian1}
\end{align}
where $\phi(\mbox{\boldmath$r$})$ is the segment density field defined as
\begin{align}
  \hat{\phi}(\mbox{\boldmath$r$})=\sum_i \delta(\mbox{\boldmath$r$}-\mbox{\boldmath$r$}^{(i)}).
\end{align}
We assume that the spatial correlation in the interaction term of the Hamiltonian (\ref{eqn:hamiltonian1}) is negligible: 
\begin{align}
  \mathcal{W} (\mbox{\boldmath$r$}, \mbox{\boldmath$r$}^\prime) \simeq \frac{v}{2}\delta (\mbox{\boldmath$r$}-\mbox{\boldmath$r$}^\prime), 
\end{align}
where $v$ is the excluded volume parameter and the prefactor $1/2$ avoids double-counting of the interaction.
Then, Hamiltonian (\ref{eqn:hamiltonian1}) becomes
\begin{align}
  \mathcal{\hat{H}}=\mathcal{\hat{H}}_{\rm ideal} + \frac{v}{2}\int d\mbox{\boldmath$r$}\left\{ \hat{\phi}(\mbox{\boldmath$r$}) \right\}^2\label{eqn:hamiltonian2}.
\end{align}
The partition function is
\begin{align}
  Z&=Z_{\rm ideal}-Z_{\rm ideal}\frac{\beta v}{2}\int d\mbox{\boldmath$r$} \left\langle \left\{ \hat{\phi}(\mbox{\boldmath$r$}; \Gamma)  \right\} ^2 \right\rangle_{\rm ideal}, \label{eqn:partition_function}
\end{align}
where $\beta v$ is regarded as the perturbation parameter.
$\langle \cdots \rangle_{\rm ideal}$ is the ensemble average value of $\cdots$ over the ensembles of the ideal system, and $Z_{\rm ideal}$ is the partition function of the ideal system.
We derive the free energy from eqn.~(\ref{eqn:partition_function}) as follows
\begin{align}
  F&\simeq -k_{\rm B}T\ln{Z_{\rm ideal}}+\frac{v}{2}\int d\mbox{\boldmath$r$} \left\langle \left\{ \hat{\phi}(\mbox{\boldmath$r$}; \Gamma)  \right\} ^2 \right\rangle_{\rm ideal}.\label{eqn:MF_free_energy}
\end{align}

By replacing $Z_{\rm ideal}$ in eqn.~(\ref{eqn:MF_free_energy}) as $Z_{\rm loop}$ and $Z_{\rm free}$, we obtain the free energy of the interacting loop and that of the unlooped chain with the excluded volume interaction, respectively.
Here, for simplicity, we assume that the segment density is spatially uniform:
\begin{align}
  \left\langle \left\{ \hat{\phi}(\mbox{\boldmath$r$}; \Gamma)  \right\} ^2 \right\rangle_{\rm ideal} \rightarrow \left[\frac{3}{4\pi} \frac{N+1}{R_{\rm g}^3} \right]^2 \label{eqn:MF},
\end{align}
where $R_{\rm g}$ is the gyration radius of the polymer chain, defined as
\begin{align}
  R_{\rm g}=\sqrt{\frac{1}{N+1}\sum_{i=0}^N \left\langle (\mbox{\boldmath$r$}^{(i)}-\mbox{\boldmath$r$}_{\rm g})^2 \right\rangle_{\rm ideal}},
\end{align}
where $\mbox{\boldmath$r$}_{\rm g}$ is the center of mass of the chain
\begin{align}
  \mbox{\boldmath$r$}_{\rm g}=\frac{1}{N+1}\sum_{i=0}^N \mbox{\boldmath$r$}^{(i)}.
\end{align}
It is noted that, as the scheme of the transfer matrix incorporates the correlation between the orientations of two consecutive segments through the chain stiffness, the calculation of the gyration radius should be included in the correlation shown in \ref{sec:gyration_radius}.
Then, we calculate the free energy $F$ as
\begin{align}
  F&=F_{\rm 0} + \frac{v_{\rm ex}^{(\rm loop)}}{2}\frac{\alpha (N+1)^2}{ \left\{ R_{\rm g}^{(\rm loop)} \right\}^3}
  -\frac{v_{\rm ex}^{(\rm free)}}{2}\frac{\mathcal{V}(N+1)^2}{ \left\{ R_{\rm g}^{(\rm free)} \right\}^6} \label{eqn:realistic_loop},
\end{align}
where $v_{\rm ex}^{(\rm loop)}, \alpha,  R_{\rm g}^{(\rm loop)}, v_{\rm ex}^{(\rm free)}$, $\mathcal{V}$, and $R_{\rm g}^{(\rm free)}$ are the excluded volume parameters of a single loop, the number of loops interacting with each other, the gyration radius of the phantom loop, the excluded volume parameter of the unlooped chain, the volume of the system, and the gyration radius of the unlooped chain, respectively.
Note that the interaction term of the looped chain in eqn.~(\ref{eqn:realistic_loop}) is obtained from the integration interval of eqn.~(\ref{eqn:MF_free_energy}) which is $\alpha \times R_{\rm g}^3$.
For simplicity, we set
\begin{align}
  v_{\rm ex}^{(\rm free)}=0.
\end{align}
We obtain the free energy of the interacting loop
\begin{align}
  F&=F_{\rm 0} + \frac{v_{\rm ex}^{(\rm loop)}}{2}\frac{\alpha N^2}{ \left\{ R_{\rm g}^{(\rm loop)} \right\}^3} \label{eqn:realistic_loop2}.
\end{align}
For the sake of simplicity, we denote $v_{\rm ex}^{\rm (loop)}$ as $v_{\rm ex}$. 
\subsection{Free energy of phantom loop}
\hspace{6mm}First, we calculate the increase in the free energy due to the loop constraint without the excluded volume interaction. 
The chromatin / DNA fiber is characterized by the chain stiffness which is described by the persistence length $l_{\rm p}$ as follows
\begin{align}
  l_{\rm p}&=\left(\frac{1}{2\delta}+\frac{3}{2} \right)b.\label{eqn:persistence_length}
\end{align}
We obtain this expression by fitting the segment orientation correlation along the chain.
The details of the fitting are shown in \ref{sec:lp}.
Here, we take $30$ nm  {as} the persistence length of chromatin\cite{chicken_chromatin_persistence_length}, although the experimental data of the persistence length on chromatin are widely distributed because of the difficulty of the measurements \cite{chicken_chromatin_persistence_length, yeast_chromatin_lp1}.
Following \cite{Goloborodko}, we set the segment size to $b=10$ nm, which specifies the minimum spatial scale of the system.
 {The Domb-Joyce model with bending elasticity} also describes bare DNA as $l_{\rm p}=50$ nm, which is the value of DNA\cite{plasmid_DNA_persistence_length}.

\begin{figure}[H]
  \begin{center}
    \includegraphics[width=12cm]{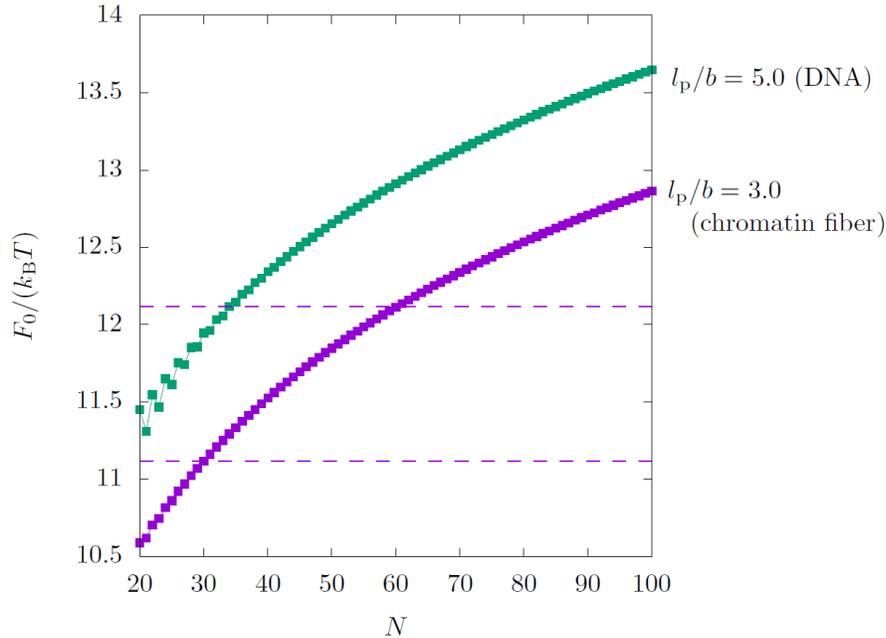}
    \caption{
    Free energy of a phantom loop.
    The vertical and horizontal axes represent the non-dimensional free energy $F_0/(k_{\rm B}T)$ and the number of segments in loop $N$, respectively.
    The purple and green symbols represent the free energies of chromatin and bare DNA, respectively.
    The persistence length $l_{\rm p}$ identifies the difference between these polymers.
    The two purple dashed lines represent the values of $F_0(N=30, l_{\rm p}/b=3.0) / (k_{\rm B}T)$ and $(F_0(N=30, l_{\rm p}/b=3.0) +k_{\rm B}T)/ (k_{\rm B}T)$.
    }
    \label{fig:ideal_loop}
 \end{center}
\end{figure}

We calculate the free energy of a single phantom loop with $l_{\rm p}=30$ nm, which corresponds to a chromatin loop.
As shown by the purple line in Fig.~\ref{fig:ideal_loop}, $F_0$ is an increasing function of $N$, and the slope decreases with $N$.
The free energy is well fitted by
\begin{align}
  \beta F_0 \simeq 1.45 \times \ln(N) + 6.15,\label{eqn:ideal_free_energy_fitting}
\end{align}
where the non-linear least squares method is used for fitting.
\subsection{Free energy of interacting loop}
\hspace{6mm}We show the increase in free energy owing to the loop constraint including the excluded volume interaction (interacting loop model).
In this  {interacting loop}, the free energy $F$ depends not only on $N$ and $l_{\rm p}$ but also on the parameters $\alpha$ and $v_{\rm ex}$. 
We adopt $\alpha = 20 $, which is estimated by the number of loops overlapping each other according to the chromosome conformation reported by Gibcus {\it et al.}\cite{Gibcus_helix} (see \ref{sec:alpha} for more details).

\begin{figure}[H]
  \begin{center}
    \includegraphics[width=12cm]{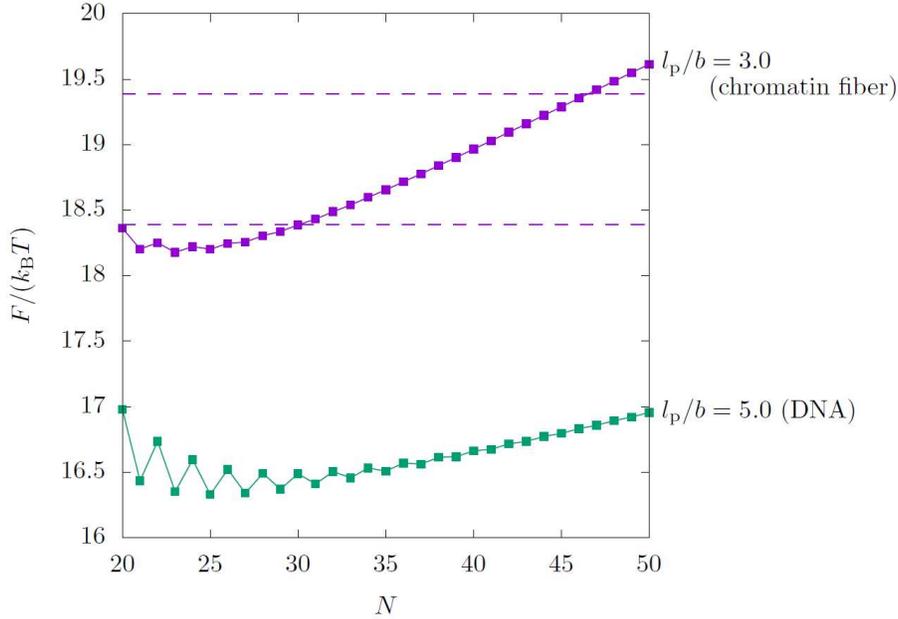}
    \caption{
   Free energy of the interacting loop considering $\alpha=20$ and $\beta b^3 v_{\rm ex} =5.0 \times 10^{-2}$.
    The vertical and horizontal axes represent the non-dimensional free energy $F/(k_{\rm B}T)$ and the number of segments in loop $N$, respectively.
    The purple and green symbols depict the free energies of the interacting chromatin loop and the interacting bare DNA loop, respectively.
    The two purple dashed lines represent the same as those in Fig.~\ref{fig:ideal_loop}.
    }
    \label{fig:realistic_loop}
 \end{center}
\end{figure}

We first calculate the free energy with $\beta b^3 v_{\rm ex}  =5.0 \times 10^{-2}$ to obtain the general trend of the free energy.
As shown in Fig.~\ref{fig:realistic_loop} (purple line), the free energy of the interacting chromatin loop increases linearly with the number of segments within a large $N$ range ($N>30$).
\begin{align}
  F =\lambda(\alpha,\beta v_{\rm ex}) N.
\end{align}
We determine the linear coefficient as follows
\begin{align}
  \lambda(\alpha,\beta v_{\rm ex}) = \left[\frac{\beta F(N=50)-\beta F(N=40)}{10} \right].\label{eqn:lambda}
\end{align}
With $\alpha = 20$ and $\beta b^3 v_{\rm ex} =5.0 \times 10^{-2}$, $\lambda(\alpha,\beta v_{\rm ex})\simeq 15$.
The free energy of DNA is shown in Fig.~\ref{fig:realistic_loop} (green lines).

It should be noted that the free energy of DNA is smaller than that of the chromatin fiber irrespective of $N$ because of the large gyration radius of DNA (see eqn.~(\ref{eqn:realistic_loop2})), originating from the large persistence length.
In the small $N$ region, the large persistence length leads to the discrete behavior of the free energy of DNA, which comes from the characteristics of the discrete lattice model.
\subsection{Typical increases of the loop lengths during the chemical reaction cycle}
We consider that a chromatin loop with length $N$ is forced to increase with the thermal-driving scenario.
In this scenario the unidirectional change in loop length is ensured by the non-equilibrium reaction coupled with ATP hydrolysis in the chemical reaction cycle and the deficit of free energy is supplied from thermal energy.
Under thermal fluctuation, a physical object typically gains $1k_{\rm B}T$ from thermal noise.
For the phantom loop case, the typical increase in the number of segments of the loop during a chemical reaction cycle (single-cycle growth in the phantom loop) $x_{\rm ph}(N)$ is calculated as follows (see the two purple dashed lines in Fig.~\ref{fig:ideal_loop}).
\begin{align}
  1.0 &= \beta F_0(N+x_{\rm ph}(N))-\beta F_0(N).\label{eqn:thermal_with_ideal}
\end{align}
Note that the contribution of the constant to $F_0$ (see eqn.~(\ref{eqn:ideal_free_energy_fitting})) disappears in eqn.~(\ref{eqn:thermal_with_ideal}).
Using eqn.~(\ref{eqn:ideal_free_energy_fitting}), we obtain
\begin{align}
  x_{\rm ph}(N)&=N\exp\left(\frac{1.0}{1.45} \right)-N\label{eqn:x_therm_ph}.
\end{align}
In the case of $N=30$, the single-cycle growth in the phantom loop is $x_{\rm ph}(30)=30$.

\begin{figure}[H]
  \begin{center}
   \includegraphics[width=12cm]{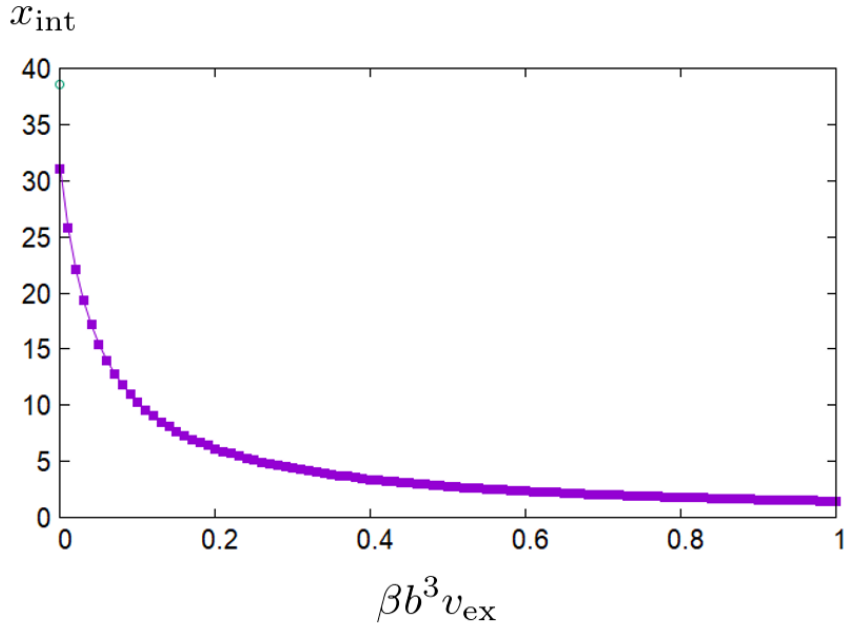}
    \caption{
    Behavior of $x_{\rm int}$ with $\beta v_{\rm ex} b^3$.
    The number of loops $\alpha = 20$ is fixed.
    The green symbol is the value of $x_{\rm ph}$, derived by substituting $N=40$ in eqn.~(\ref{eqn:x_therm_ph}).
    }
    \label{fig:growth_rate_thermal}
 \end{center}
\end{figure}

As the candidate for the resistance force is the excluded volume interaction, we demonstrated the typical increase in the number of segments in a single chemical reaction cycle based on  {an} interacting loop (single-cycle growth in the interacting loop) with excluded volume parameter.
Then, we consider the single-cycle growth for the interacting loop, whose growth is resisted by the excluded volume interaction.
As the free energy behaves as a linear function of $N$, the single-cycle growth in the interacting loop, $x_{\rm int}$,  is estimated from the slope of the free energy.
Here, $x_{\rm int}$ is defined as 
\begin{align}
 x_{\rm int}=\lambda^{-1}(\alpha,\beta v_{\rm ex})\label{eqn:def_of_x_therm}.
\end{align}
The single-cycle growth in the interacting loop depends on the excluded volume parameter $\beta v_{\rm ex}$, as shown in Fig.~\ref{fig:growth_rate_thermal}.
Note that $x_{\rm int} (\beta b^3 v_{\rm ex} =0)$ deviates from $x_{\rm ph}$ because the definitions of these (eqns.~(\ref{eqn:x_therm_ph}) and (\ref{eqn:def_of_x_therm})) are different from each other.

At $\beta b^3 v_{\rm ex} =0$, as shown in Fig.~\ref{fig:growth_rate_thermal}; because there is a small difference between $x_{\rm ph}$ and $x_{\rm int}$, the interacting loop may be connected to the phantom loop.
With a large $\beta b^3 v_{\rm ex}$, the difference between $x_{\rm ph}$ and $x_{\rm int}$ becomes remarkably large.
This means that the excluded volume interaction is important for quantifying single-cycle growth.
We therefore conclude that the phantom loop is not  {appropriate} for quantification of the chromatin loop in the chromosome.
\subsection{Loop growth with iterative reaction cycles}\label{sec:loop_length_model}
The cyclic chemical reaction drives the growth of the chromatin loop length (loop growth), including the ATP hydrolysis of condensin.
In this subsection, we construct  {an iterative map} that describes the loop growth dynamics for the interacting loop as the reaction cycle proceeds.
 {The} iterative map represents the model $N_{i+1}=\mathcal{F}(N_i)$, where   {$N_i$} is the length of the loop  {at} the iteration number of the chemical reaction cycles  {$i$} .
Although the iterative map for the phantom loop is identical to that in eqn.~(\ref{eqn:x_therm_ph}), the loop growth for the interacting loop increases the nucleosome density, which results in non-trivial changes in the excluded volume interaction.
First, we outline the construction of the iterative map. 
We obtain the relationship between the excluded volume parameter and the density $v_{\rm ex}=g(\rho)$.
Moreover, we assume the relationship between the loop length and nucleosome density $\rho = f(N)$ .  
Then, we construct the map and investigate the loop growth dynamics.
Because the definition range of $\rho$ is determined by the experimental results, we introduce the actual spatial scale.
Therefore, the loop length (in the actual scale) is defined as $L=Nb$ $( N \times 10 [{\rm nm}])$ and single-cycle growth (in the actual scale) as $l=x_{\rm int}b$ $(x_{\rm int}  \times 10 [{\rm nm}] )$.
\subsubsection{Iterative process for computing loop length}\label{sec:loop_length_model}
The cyclic reaction of growing the chromatin loop includes simultaneous changes of variables: the loop length $L$, the excluded volume parameter $v_{\rm ex}$ and the nucleosome density $\rho$ , all which regulate each other.
Here, we outline the construction of an iterative map to clarify the relationship among these variables and introduce the updating order for each. 

At the beginning of the $i$-th reaction cycle, we have the loop length $L_{i}$ and the nucleosome density $\rho_{i}$.
Then we calculated the next reaction cycle as:
\begin{enumerate}
  \item We also evaluate the excluded volume parameter from the nucleosome density $v_{\rm ex} (\rho_{i})$ .
  \item We calculate the free energy by substituting $v_{\rm ex} (\rho_{i})$ into eqn.~(\ref{eqn:realistic_loop2}).
  \item We calculate the single-cycle growth $l (\rho_{i})$ from eqn.~(\ref{eqn:def_of_x_therm}).
   The loop length then becomes $L_{i+1}=L_{i} + l (\rho_{i})$.
  \item We update the nucleosome density using the present loop length $\rho_{i+1}=\rho (L_{i+1})$.
\end{enumerate}
The ATP hydrolysis rate of the condensin is experimentally measured as $k_{\rm cycle}=0.90\ [ {{\rm s}}^{-1}]$\cite{DNA_supercoil_induced_by_condensin}.
Therefore, the typical time scale of the chemical reaction cycle is $1/0.90\simeq 1.11$ [s].
As a chromosome matures at 3600 [s]\cite{Gibcus_helix}, the average number of reaction cycles is $3600\times 0.90=$3240.
Therefore, it takes 3240 iterations of these procedures for maturing a chromosome.

In the above iteration procedure, we use two functions, $v_{\rm ex} (\rho)$ and $\rho (L)$, which are undefined.
These functions are introduced in the next section.
\subsubsection{Dependence of excluded volume parameter on nucleosome density and dependence of nucleosome density on loop length}
The excluded volume parameter, $v_{\rm ex}$, describes the \lq\lq{}mean (potential) field\rq\rq{}.
In mean-field theory,  mean (potential) field describes the nucleosome density around the chromatin loop that we focus on.
Thus, $v_{\rm ex}$ depends on the nucleosome density around the chromatin loop.
We estimated the excluded volume interaction with the Carnahan-Starling equation of state, which is derived from the Virial expansion, including the higher order of the system composed of rigid spheres\cite{hansen}. 
We assume $v_{\rm ex}$ to be
\begin{align}
  \beta v_{\rm ex}&=\frac{4\eta -3\eta^2}{(1-\eta)^2}, \label{eqn:CH}
\end{align}
where $\eta =\pi \rho/6$ is the packing ratio, and $\rho$ is the nucleosome density.
We show the detailed calculation in \ref{sec:CS}.
The initial and final values of $\rho$ are $0.030 \times 10^{-3} {\rm [/nm^3]}$ and $0.59\times 10^{-3} {\rm [/nm^3]}$, respectively, where the details of the calculations are shown in \ref{sec:alpha}. 

Because loop extrusion drives chromosome condensation, the nucleosome density $\rho$ surrounding the loop should be a monotonically increasing function of the length of the loop.
 Here we arrange two examples of $\rho(L)$, 
\begin{align}
  \rho_{\rm lin}(L)&=\left( \frac{0.56}{24000{\rm [nm]}}L{\rm [nm]}+0.030 \right)\times 10^{-3} {\rm [/nm^3]}\label{eqn:linear}\\
  \rho_{\rm exp}(L)&=\left(\frac{0.56}{e-1}\exp\left[\frac{L{\rm [nm]}}{24000{\rm [nm]}}\right]+0.030-\frac{0.56}{e-1}\right)\times 10^{-3} {\rm [/nm^3]} \label{eqn:exp}
\end{align}
where both functions satisfy $\rho(L=0{\rm [nm]})=0.030 \times 10^{-3} {\rm [/nm^3]}$ and $\rho(L=24000{\rm [nm]})=0.59\times 10^{-3} {\rm [/nm^3]}$.
The former function assumes that the nucleosome density linearly depends on the loop length, while the latter assumes that the nucleosome density exponentially increases with the loop length.
\subsubsection{Chromatin loop length with the number of reaction cycles}
Using the procedure described in subsection.~\ref{sec:loop_length_model}  using eqns.~(\ref{eqn:CH}), (\ref{eqn:linear}), and (\ref{eqn:exp}) , we calculated the loop growth dynamics with the iteration of the reaction cycles.
\begin{figure}[H]
  \begin{center}
   \includegraphics[width=12cm]{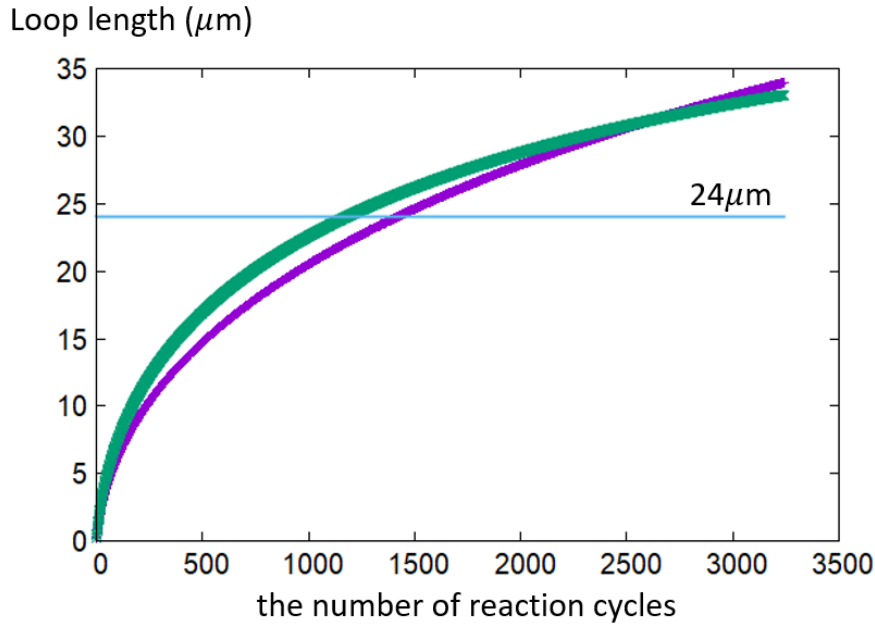}
    \caption{
    Loop length on the number of reaction cycles with eqn.~(\ref{eqn:linear}) (purple) and eqn.~(\ref{eqn:exp}) (green).
    The vertical and horizontal axes represent the loop length and the number of reaction cycles, respectively.
    The cyan line shows $24\mu$m, which corresponds to the chromatin loop length in the mature chromosome.
    }
    \label{fig:loop_length}
 \end{center}
\end{figure}
Figure \ref{fig:loop_length} shows that the loop length reaches the matured chromosome.
Therefore, this result implies that the thermal driving force under the assumption of eqn.~(\ref{eqn:linear}) or (\ref{eqn:exp}) is sufficient to mature the chromosome.
Under the assumption of eqn.~(\ref{eqn:linear}), the mature chromosome is constructed at approximately 1500 cycles, whereas with eqn.~(\ref{eqn:exp}), 1000 cycles is approximately required for the matured chromosome.
\section{Discussion and conclusion}
\hspace{6mm}In our study, we employed the lattice polymer model  {(Domb-Joyce model with bending elasticity)} to describe a chromatin fiber and determine the free energy of the looped state.
We incorporated the excluded volume interaction using the mean-field theory, where the excluded volume parameter $v_{\rm ex}$ represented the strength of the interaction.
We found that the free energy in the phantom loop depends logarithmically on the loop length, whereas the free energy in the interacting loop was proportional to the loop length.
We also evaluated the single-cycle growth dependence on the excluded volume parameter, which reveals how  {the} chromosome condensation suppresses the growth of the chromatin loop. 

\begin{figure}[H]
  \begin{center}
   \includegraphics[width=12cm]{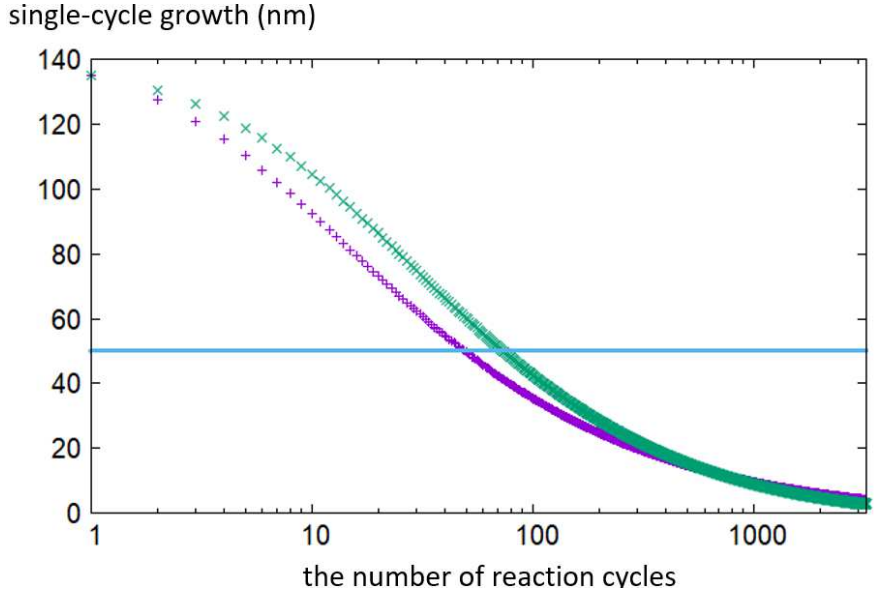}
    \caption{
    The dependence of single-cycle growth on the number of reaction cycles with eqn.~(\ref{eqn:linear}) (purple) and eqn.~(\ref{eqn:exp}) (green).
    The vertical and horizontal axes represent the single growth cycle and the number of reaction cycles, respectively.
    The cyan line shows 50 nm which corresponds to the size of condensin.
    }
    \label{fig:x_with_time}
 \end{center}
\end{figure}

We constructed an iterative map to describe the growth of the chromatin loop length, along with the ATP-hydrolytic reaction cycle of condensin.
By assuming the dependence of $v_{\rm ex}$ on $\rho$ based on the Carnahan-Starling equation of state, we obtained the loop growth dynamics along with the reaction cycles.
We found the thermal fluctuation to be sufficient as an energy source to grow the chromatin loop and to mature the chromosome.
We assumed the excluded volume parameter to be a monotonically increasing function of the loop length, such as eqns.~(\ref{eqn:linear}) and (\ref{eqn:exp}).
The loop length is assumed to increase with the number of reaction cycles.
Due to these assumptions, the single-cycle growth $l$ [nm] is a decreasing function of the number of reaction cycles, as shown in Fig.~\ref{fig:x_with_time}.
The condition of $l= 50$ nm corresponds to $\beta v_{\rm ex} \simeq 0.25 \times 10^{-3} [{\rm nm}^{-3}] \ (\rho \simeq  0.11 \times 10^{-3} [{\rm nm}^{-3}])$. This small value means that the perturbation expansion can describe this condition regarding the density up to the second order.
We also confirmed the validity of the perturbation theory using another  {map} in \ref{sec:modified_interaction}.
 {\ref{sec:modified_interaction} represents the  {map which incorporates the} free energy directly incorporating the higher-order terms of the perturbation.}
 {The  {result based on the map} in \ref{sec:modified_interaction} qualitatively exhibits the same result as Fig.~\ref{fig:x_with_time}.}

\begin{figure}[H]
  \begin{center}
    \includegraphics[width=12cm]{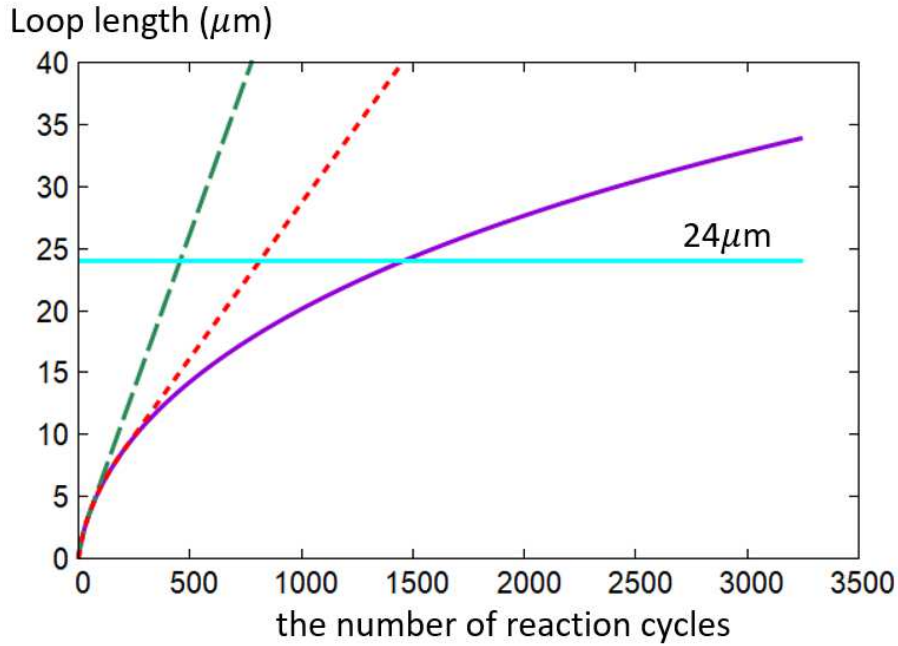}
    \caption{
    Dependence of the loop length on the number of reaction cycles with eqn.~(\ref{eqn:linear}).
    The purple solid line represents the loop length with the number of reaction cycles, which is the same as that in Fig.~\ref{fig:loop_length}.
    The green dashed and red dotted lines show that the single-cycle growth becomes constant when it reaches 50 nm (green dashed line) or 25 nm (red dotted line).
    The cyan line represents 24 $\mu$m, which corresponds to the chromatin loop length in the mature chromosome.
    }
    \label{fig:modified_loop_length_linear}
 \end{center}
\end{figure}

\begin{figure}[H]
  \begin{center}
    \includegraphics[width=12cm]{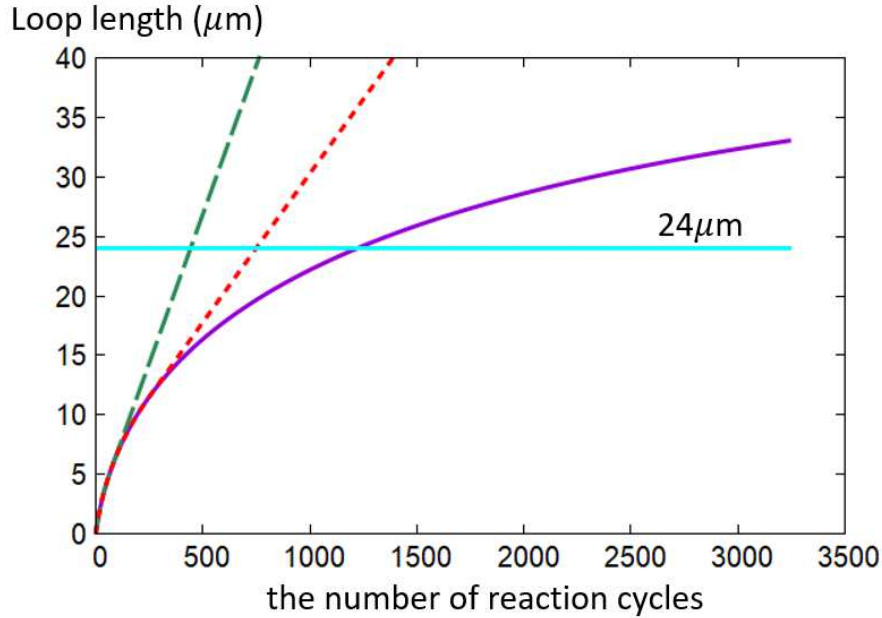}
    \caption{
    Dependence of the loop length on the number of reaction cycles when eqn.~(\ref{eqn:exp}) is assumed.
    The purple solid line represents the loop length with the number of reaction cycles, which is the same as that in Fig.~\ref{fig:loop_length}.
    The green dashed and red dotted lines are the same as those shown in Fig.~\ref{fig:modified_loop_length_linear}.
    The cyan line represents 24 $\mu$m, which corresponds to the chromatin loop length in the mature chromosome.
    }
    \label{fig:modified_loop_length_exp}
 \end{center}
\end{figure}

In the calculation, we did not consider the size, shape, and detailed movement of the condensin.
However, Fig.~\ref{fig:x_with_time} shows that the single-cycle growth can be smaller than the size of the condensin ($\sim 50$ nm).
Therefore, we should consider the interference in the behavior when they have a similar size.
Moreover, we should also discuss the behavior of condensin and  chromatin loop captured by it.
The  {detailed processes} of the thermal-driving scenario involve chromatin capture by condensin and conformational changes in chromatin and condensin.
The condensin has two coiled-coil arms of $50$ nm and is considered to capture DNA at multiple domains, including the hinge and the head of the arms. 
If the essential steps in the reaction cycle of loop extrusion include DNA captured simultaneously by these two domains (hinge and head), the length between the captured sites along DNA must be longer than the distance between the domains.  
This shows that the minimum value of single-cycle growth can be restricted by the distance between the hinge and head domains.
 {After the single-cycle growth drops to the condensin size, it should not decrease further; i.e., the constant single-cycle growth comparable to the condensin size. 
This implies the direct power stroke of condensin as a molecular motor for loop extrusion, which can be called  as a motor-pulling scenario.}
Moreover, the loop extrusion can be regarded as a switching of two processes; the former is carried by the thermal-driving scenario and the latter is carried by the motor-pulling scenario.
Suppose the distance between the hinge and head domains to be 50 nm or 25 nm and employing the switching scenario, the loop growth dynamics are modified as shown in Figs.~\ref{fig:modified_loop_length_linear} and \ref{fig:modified_loop_length_exp}.
\hspace{-3mm}Until the single-cycle growth reaches the restricted value (50 or 25 nm), the loop growth dynamics remain the same as that in Fig.~\ref{fig:x_with_time}, and then the loop length grows linearly after that.
Consequently, the loop length reaches a mature value earlier than only the thermal-driving scenario. 

 {In this study, the lattice polymer model is employed to calculate the conformational free energy of the looped semiflexible chain, and the mean field approximation is employed to incorporate the excluded volume effect.} 
 {These results presented in Figs.~\ref{fig:ideal_loop} and \ref{fig:realistic_loop} are qualitatively supported by }  {Domb-Joyce model which is a lattice model for a flexible chain.}
 {The free energy of the looped chain is proportional to the logarithm of the loop length as illustrated in Fig.~\ref{fig:ideal_loop}, which is consistent with the result in the Domb-Joyce model without the excluded volume effect\cite{RW_lattice}.}
 {In \cite{DJ, RW_lattice}, the excluded volume effect is incorporated by exactly counting the number of conformations with collisions of segments.}
 {Furthermore, in this case , the dominant term of the free energy of the looped chain linearly increases with the chain length, which is consistent with the result shown in Fig.~\ref{fig:realistic_loop}. }
 {The chain stiffness results in the quantitative differences between  {the loop} free energies of the Domb-Joyce model}  {with/without the excluded volume effect and the phantom/interacting loop free energies.}
 {When}  {the}  {phantom loop is modified to the flexible chain ($\mathcal{T}_{\mu\nu} =1$ independent of $\mu$ and $\nu$),  {this} is completely the same as the  {loop conformation in} Domb-Joyce model without the excluded volume effect on the fcc lattice.}
 {Both free energies are}  {proportional to $1.5 \times \ln{N}$.}

We evaluated the single-cycle growth based on the free energy of the chromatin loop formation, which corresponds to the maximum mechanical work for loop formation.
Moreover, in the thermal-driving scenario, the slippage of the condensin along the chromatin is ignored.
Therefore, the single-cycle growth (Figs.~\ref{fig:growth_rate_thermal} and \ref{fig:x_with_time}) and loop length (Fig.~\ref{fig:loop_length}) are the upper bounds of these values.
The actual loop length with the number of reaction cycles  {could} be smaller than that shown in Fig.~\ref{fig:loop_length}.
The single-cycle growth in vivo might also be smaller than that shown in Figs.~\ref{fig:growth_rate_thermal} and \ref{fig:x_with_time}.
Therefore, the number of reaction cycles in which the thermal-driving scenario is connected to the motor-pulling scenario might be more than  {that of} our evaluation.
\appendix
\def\thesection{Appendix \Alph{section}}
\section{Gyration radius}\label{sec:gyration_radius}
The square of the gyration radius $R_{\rm g}^2$ is defined as 
\begin{align}
  R_{\rm g}^2&=\frac{1}{N+1}
  \sum_{i=0}^{N+1}\left\langle \left(\mbox{\boldmath$r$}^{(i)}-\mbox{\boldmath$r$}_{\rm g} \right)^2 \right\rangle \label{eqn:gyration_radius},
\end{align}
where $\mbox{\boldmath$r$}_{\rm g}$ is the center of mass: 
\begin{align}
  \mbox{\boldmath$r$}_{\rm g}&=\frac{1}{N+1}\sum_{j=0}^{N}\mbox{\boldmath$r$}^{(j)}.
\end{align}
Equation~(\ref{eqn:gyration_radius}) is rewritten as
\begin{align}
   R_{\rm g}^2&=\frac{1}{N+1}
  \sum_{i=0}^{N}\left\langle \left(\mbox{\boldmath$r$}^{(i)}-\mbox{\boldmath$r$}_{\rm g} \right)^2 \right\rangle \\
  &=\frac{1}{N+1}
  \sum_{i=0}^{N}\left\langle \left(\mbox{\boldmath$r$}^{(i)}
  -\frac{1}{N+1}\sum_{j=0}^{N}\mbox{\boldmath$r$}^{(j)} \right)^2 \right\rangle\\
  &=\frac{1}{N+1}\frac{1}{(N+1)^2}
  \sum_{i=0}^{N}\left\langle \left((N+1)\mbox{\boldmath$r$}^{(i)}
  -\sum_{j=0}^{N}\mbox{\boldmath$r$}^{(j)} \right)^2 \right\rangle\\
  &=\frac{1}{N+1}\frac{1}{(N+1)^2}
  \sum_{i=0}^{N}\left\langle \left(
  \sum_{j=0}^{N}(\mbox{\boldmath$r$}^{(i)}-\mbox{\boldmath$r$}^{(j)}) \right)^2 \right\rangle\\
  &=\frac{1}{N+1}\frac{1}{(N+1)^2}
  \sum_{i=0}^{N}\left\langle\sum_{j=0}^{N} \left(
  \mbox{\boldmath$r$}^{(i)}-\mbox{\boldmath$r$}^{(j)} \right)^2  
  +\sum_{j=0}^{N}\sum_{k\neq j} 
  (\mbox{\boldmath$r$}^{(i)}-\mbox{\boldmath$r$}^{(j)}) \cdot (\mbox{\boldmath$r$}^{(i)}-\mbox{\boldmath$r$}^{(k)})  \right\rangle \label{eqn:Rg_separate}.
\end{align}
\section{Determining of expression of persistence length}\label{sec:lp}
\begin{figure}[H]
  \begin{center}
    \includegraphics[width=12cm]{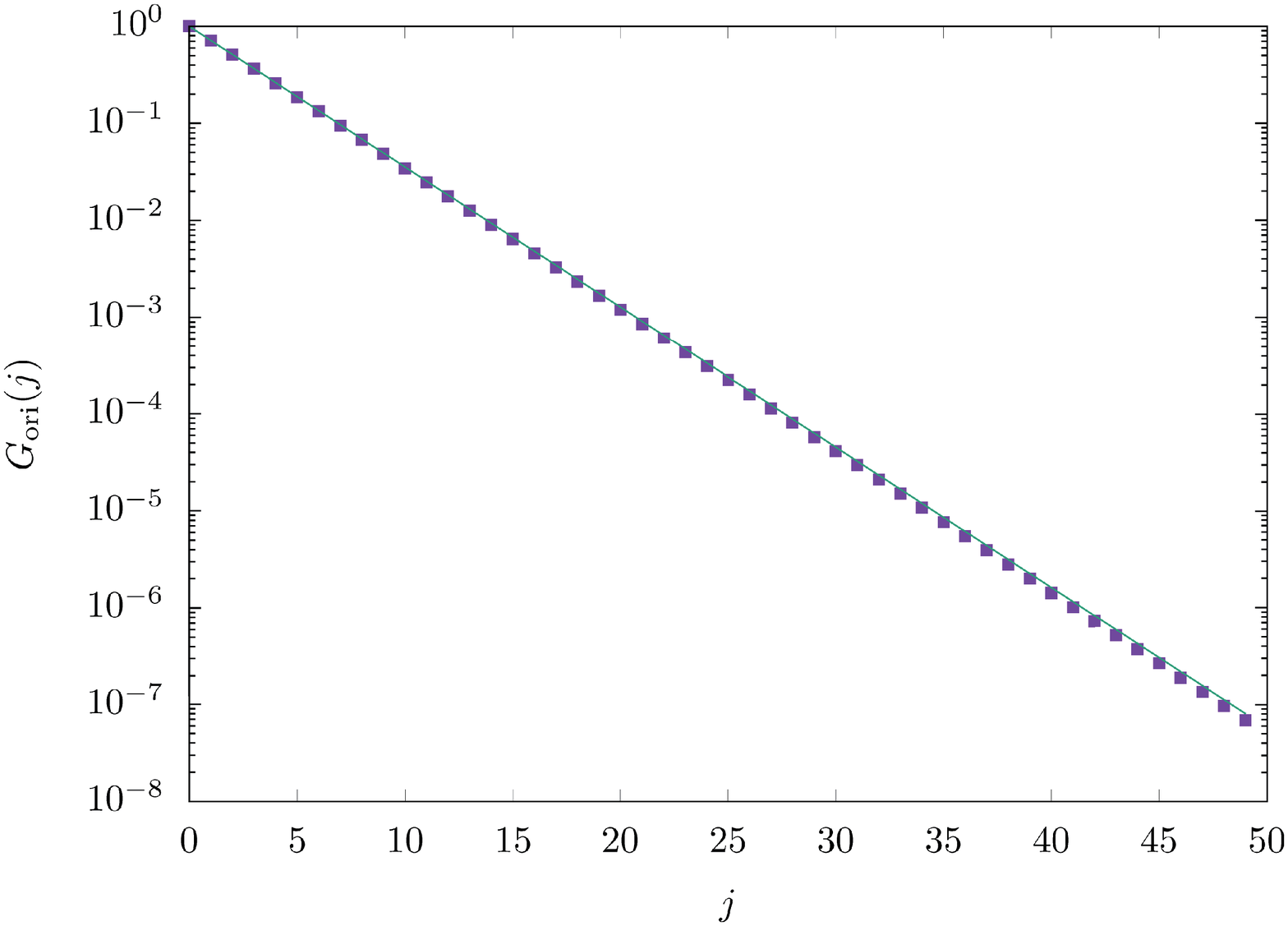}
    \caption{
    Behavior of $G_{\rm ori} (j)$ in the case of $\delta^{-1}=3.00$ (corresponding to $l_{\rm p}/b=3.00$).
    The purple and green lines represent the numerical results and the fitting function~(\ref{eqn:fitting}), respectively. 
    }
    \label{fig:orientaion_correlation_lp3}
 \end{center}
\end{figure}

\begin{figure}[H]
  \begin{center}
    \includegraphics[width=12cm]{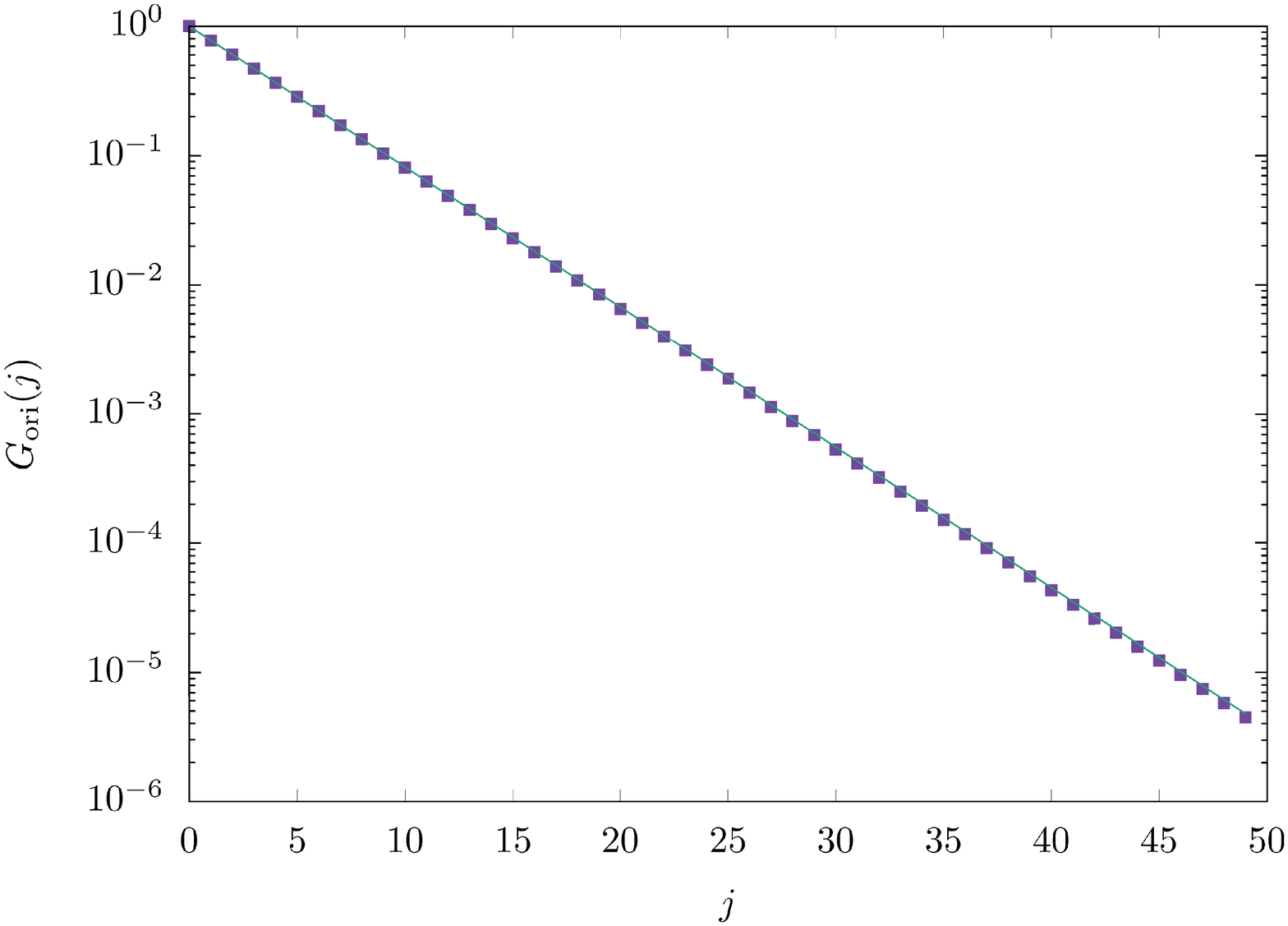}
    \caption{
    Behavior of $G_{\rm ori} (j)$ in the case of $\delta^{-1}=8.00$ (corresponding to $l_{\rm p}/b=5.00$).
    The purple and green lines are the same as in Fig.~\ref{fig:orientaion_correlation_lp3}. 
    }
    \label{fig:orientaion_correlation_lp5}
 \end{center}
\end{figure}

\hspace{3mm}To determine the expression, we compute the orientation correlation along the chain as the persistence length is defined by the contour length of the polymer chain when the segment orientation correlation is lost.
We describe the orientation correlation function $G_{\rm ori}(j)$ as
\begin{align}
  G_{\rm ori}(j)&=\langle \mbox{\boldmath$b$}^{(0)} \cdot \mbox{\boldmath$b$}^{(j)} \rangle\\
  &=\sum_{p=x, y, z}\sum_{\mu,\nu,\eta}^{12}e_{p}^{(\mu)} {\mathcal{T}^j}_{\mu\nu} e_p^{(\nu)}{\mathcal{T}^{N-j}}_{\nu\eta},
\end{align} 
where $j$ is the segment index and $p$ denotes the Cartesian coordinates $x, y$, or $z$.
The persistence length satisfies:
\begin{align}
  G_{\rm ori}(j)&=\exp\left[-\frac{j}{(l_{\rm p} / b)} \right]\label{eqn:orientation_correlation_and_persistence_length}.
\end{align}
By using the transfer matrix, we calculate $G_{\rm ori}(j)$ with $\delta^{-1}=3.00$ and $\delta^{-1}=8.00$ , as shown in Figures~\ref{fig:orientaion_correlation_lp3} and \ref{fig:orientaion_correlation_lp5}, 
We fit these numerical results by
\begin{align}
  G_{\rm ori}(j)&=\exp\left[-\frac{j}{\displaystyle \frac{\delta^{-1}+3.0}{2}} \right].\label{eqn:fitting}
\end{align}
By using equations~(\ref{eqn:orientation_correlation_and_persistence_length}) and (\ref{eqn:fitting}), we describe the persistence length $l_{\rm p}$ as
\begin{align}
  l_{\rm p}&=\left(\frac{1}{2\delta}+\frac{3}{2} \right)b\label{eqn:persistence_length_clear}.
\end{align}
When the value of $b$ is not strictly determined, the constant contribution to $l_{\rm p}$ can be renormalized in segment size $b$ (for example, where the value of $b$ simply specifies the unit length of the system \cite{Yokota_induction}).
Therefore, in such a case, the persistence length can be defined as: 
$l_{\rm p}= b / \delta$.
However, in this study, we cannot do so because the value of $b$ determines the actual spatial scale, {\it i.e.}, $b=10$[nm].
Here, we use eqn.~(\ref{eqn:persistence_length_clear}) to define the persistence length
\section{Estimation of $\alpha$}\label{sec:alpha}
\begin{figure}[H]
  \begin{center}
    \includegraphics[width=12cm]{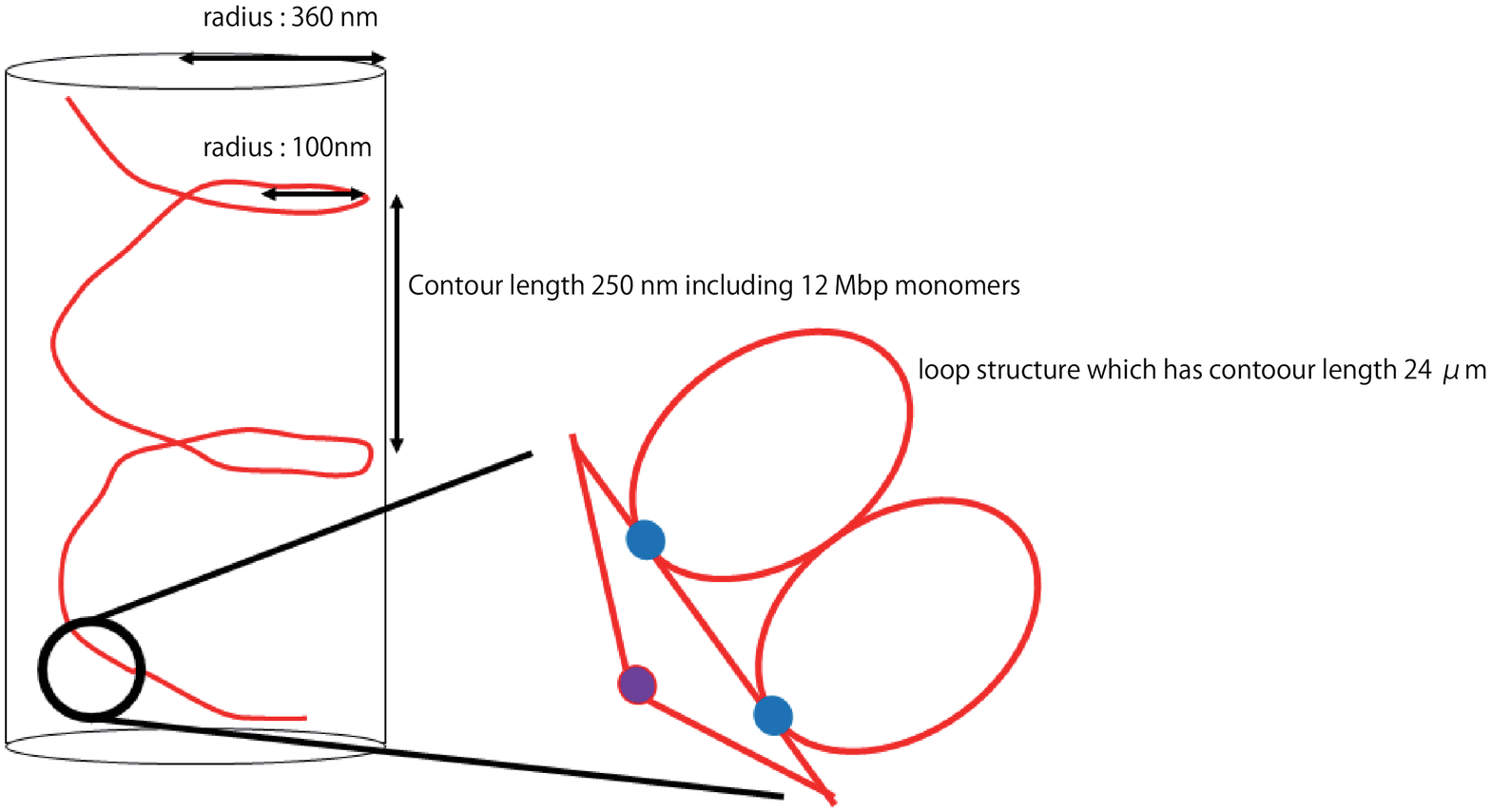}
    \caption{
    Chromosome structure reported by Gibcus {\it et al.}\cite{Gibcus_helix}, based on Hi-C data and simulation studies.
    The red line represents a chromosome with a helical conformation, including a nested loop structure.
   The blue and purple circles depict condensin I and II, respectively. 
    }
    \label{fig:chromosome_structure_Gibcus}
 \end{center}
\end{figure}
We estimate the number of loops interacting with each other $\alpha$ based on the chromatin conformation shown in Fig.~\ref{fig:chromosome_structure_Gibcus}.
We denote the distance between the centers of mass of the nearest-neighbor loops by $\Delta$.
As the gyration radius of the loop is $R_{\rm g}$, 
\begin{align}
  \alpha=\frac{R_{\rm g}}{\Delta}.
\end{align}
We calculate the number of loops, including the spatial range $\Delta$.
During prometaphase, the length of each loop is $80$ kbp.
Additionally, 1 bp $\simeq$ 0.3 nm which leads to
\begin{align}
  R_{\rm g}\sim \sqrt{N} \times {\rm size\ of\ base\ pair}=\sqrt{80\times 10^3} \times 0.3 [\rm nm]=60\sqrt{2} [\rm nm].
\end{align}
As 150 loops exist in one pitch of the helical conformation, as shown in Fig.~\ref{fig:chromosome_structure_Gibcus}.
\begin{align}
  \Delta = \frac{2\pi r_{\rm spiral}}{150}=\frac{2\pi \times 100 [\rm nm]}{150}\sim 4 [\rm nm].
\end{align}
We calculated the number of interacting loops $\alpha$ as
\begin{align}
  \alpha=\frac{R_{\rm g}}{\Delta} \simeq 20.257.
\end{align}
Therefore, we fixed $\alpha=20$.%

Next, we estimate the number of nucleosomes in the pitch (1 turn) of the helix.
\begin{align}
  \frac{12 {\rm Mbp}}{200 {\rm bp}}&=\frac{1200 \times 10^4 {\rm bp}}{200 {\rm bp}}\\
  &=6.0 \times 10^4 {\rm nucleosomes}/{\rm turn}
\end{align}
The volume of 1 turn $V$ is
\begin{align}
  V&=\pi \times 360 {\rm nm} \times  360 {\rm nm} \times 250 {\rm nm} / {\rm turn}\\
  &=\pi \times 36 {\rm segment} \times  36 {\rm segment} \times 25 {\rm segment}/ {\rm turn},
\end{align}
Then, the number of nucleosomes per $b^3$, $\rho$, is
\begin{align}
  \rho=\frac{6.0 \times 10^4 {\rm nucleosomes}/{\rm turn}}{\pi \times 36 {\rm segment} \times  36 {\rm segment} \times 25 {\rm segment}/ {\rm turn}}&\simeq 0.59 \  {\rm nucleosomes}/b^3.
\end{align}

We then evaluate the initial value of  {the} nucleosome density by referring \cite{Sakai2018}.
According to \cite{Sakai2018}, which is the study of chromosome formation using a bead-spring model, the bead density before chromosome formation is 0.010.
Although a rigorous value is not mentioned in \cite{Sakai2018}, several nucleosomes constructs a bead. 
In \cite{Sakai2018}, the bead-spring model without bending elasticity is used.
Then, we chose the initial value of the nucleosome density as 0.030 $ {\rm nucleosomes}/b^3$, meaning that the three nucleosomes (corresponding to the persistence length of the chromatin fiber, 30 nm) compose a bead. 
\section{Dependence of excluded volume parameter on density}\label{sec:CS}
In this appendix, we obtain the dependence of $\beta v_{\rm ex}$ on the nucleosome density $\rho$ from the Carnahan-Starling equation of state\cite{hansen}.
The Carnahan-Starling equation of state is that of a system composed of hard spheres, which shows good agreement with experiments even in the dense state.
In a system composed of hard spheres, the virial expansion up to the 10th order was analytically or numerically calculated.
By using the packing ratio $\eta = \pi \rho b^3/6$, we described the equation of state as
\begin{align}
  \frac{\beta P}{\rho}=1+\sum_{i=1}^{10} \mathcal{B}_i \eta^i,
\end{align}
where $P$ is the pressure of the system.
Here, $\mathcal{B}_i$ is roughly fitted as
\begin{align}
  \mathcal{B}_i=i^2+3i.
\end{align}
By using this series, the general form of the Virial expansion is interpolated as
\begin{align}
  \frac{\beta P}{\rho}&=1+\sum_{i=1}^\infty (i^2+3i) \eta^i \label{eqn:CS_original}.
\end{align}
This equation of state is the so-called Carnahan-Starling equation of state.
The first term of eqn.~(\ref{eqn:CS_original}) represents the equation of the state of the ideal system, and the second term represents the interaction between particles. 
The interaction term of the free energy $F_{\rm inter}$ is calculated by integrating the second term in eqn.~(\ref{eqn:CS_original}) with respect to the volume.
\begin{align}
  F_{\rm inter} &= - \int dV \frac{\rho}{\beta} \frac{4\eta - 2 \eta^2}{(1-\eta)^3}\\
  &=   \int_0^\eta d\eta \frac{\rho}{\beta} \frac{N}{\rho \eta} \frac{4\eta - 2 \eta^2}{(1-\eta)^3}\\
  &= N\int_0^\eta d\eta  \frac{1}{\eta} \frac{4\eta - 2 \eta^2}{(1-\eta)^3}\\
  &=N\frac{4\eta -3\eta^2}{(1-\eta)^2} .
\end{align}
As a segment in the interacting loop undergoes interaction energy $\beta v_{\rm ex}$, it can be calculated as: 
\begin{align}
  \beta v_{\rm ex} = \frac{\beta F_{\rm inter}}{N} =\frac{4\eta -3\eta^2}{(1-\eta)^2}.
\end{align}
\section{Modification of interaction term in free energy}\label{sec:modified_interaction}
In our  {iterative map} shown in the main text, the excluded volume parameter is evaluated as the interaction term of the free energy per segment (or particle) incorporating the higher-order terms of the Virial expansion.
However, $\beta v_{\rm ex}$ is the coefficient of the 2nd order term of the Virial expansion.
Moreover, in our interacting loop, we regard the number of loops interacting with each other, $\alpha$, as a constant parameter, although $\alpha$ should also depend on the nucleosome density.
Therefore, there is room for modification of the treatments of the interaction term.
In the considerable modification, we directly incorporate the contribution from the higher-order terms of the Virial expansion into the free energy: 
\begin{align}
  F=F_0 + N\frac{4\eta -3\eta^2}{(1-\eta)^2}. \label{eqn:free_energy_direct}
\end{align}
Using this expression, we can execute the same computation for the loop length as the number of reaction cycles.
In this modified  {map}, we obtain the same result quantitatively as our original  {map}, where the connection of the thermal-driving scenario to the motor-pulling scenario appears before the matured chromosome is constructed.
\begin{figure}[H]
  \begin{center}
   \includegraphics[width=12cm]{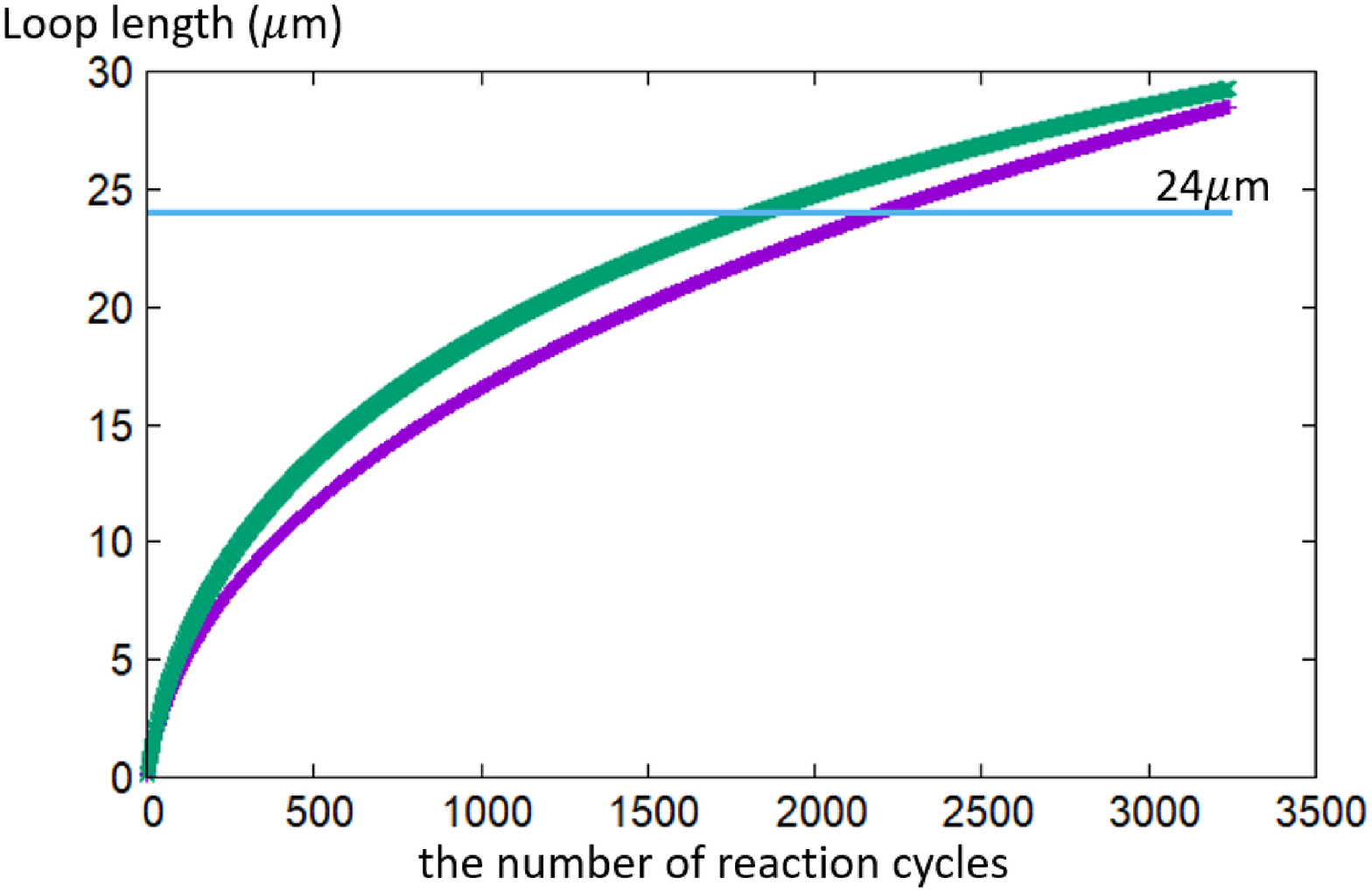}
    \caption{
    Dependence of the loop length on the number of reaction cycles based on eqn.~(\ref{eqn:free_energy_direct}).
    The vertical and horizontal axes represent the loop length and the number of reaction cycles, respectively.
    We assumed the dependence of $\rho$ on the loop length to be eqn.~(\ref{eqn:linear}) (purple) and eqn.~(\ref{eqn:exp}) (green).
    The cyan line shows 24 $\mu$m, which corresponds to the chromatin loop length in the mature chromosome.
    }
    \label{fig:loop_length_direct}
 \end{center}
\end{figure}
\begin{figure}[H]
  \begin{center}
   \includegraphics[width=12cm]{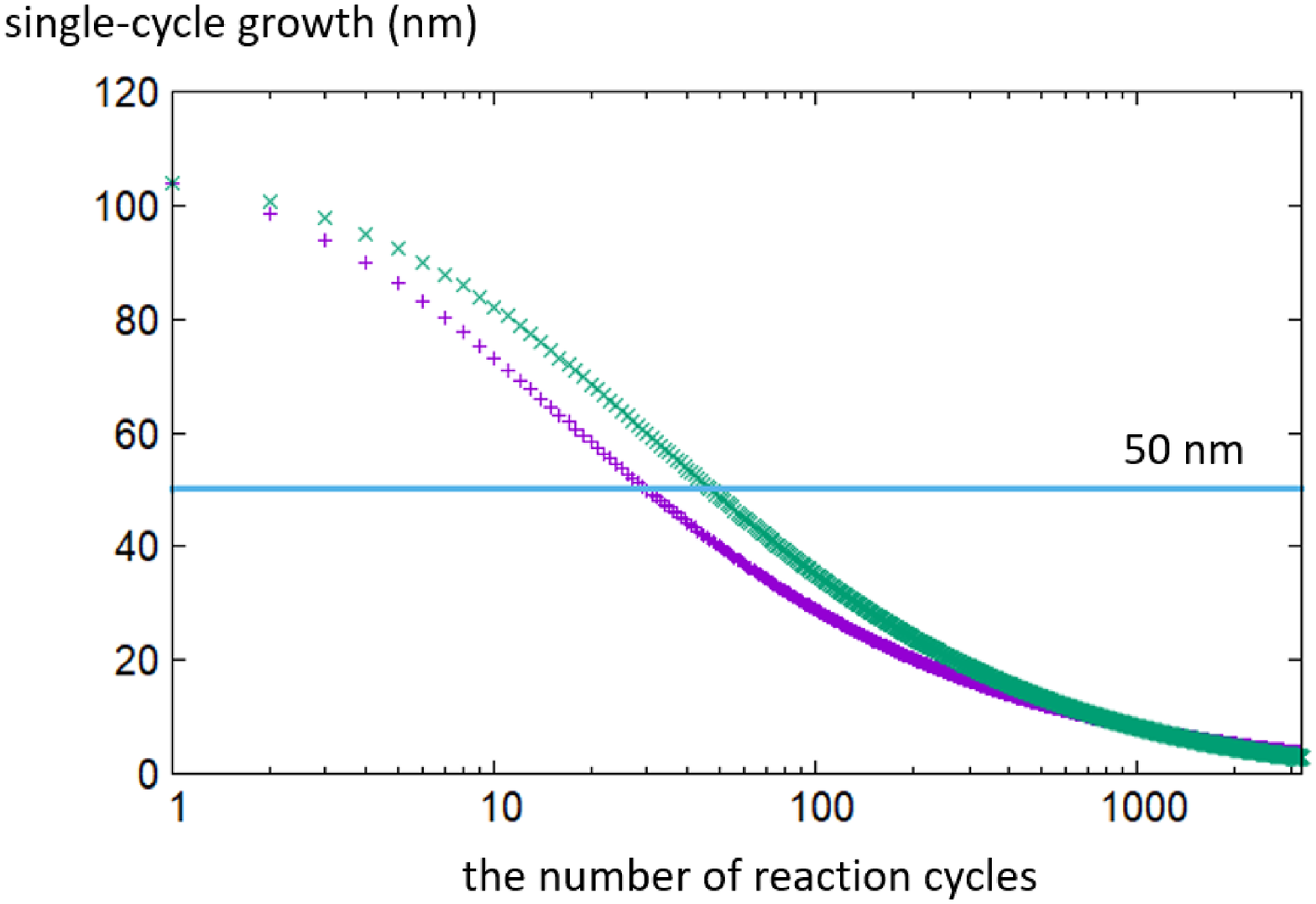}
    \caption{
    Dependence of the single-cycle growth on the number of reaction cycles based on eqn.~(\ref{eqn:free_energy_direct}).
    The vertical and horizontal axes represent the single-cycle growth and the number of reaction cycles, respectively.
    The dependence of $\rho$ on the loop length is assumed to be eqn.~(\ref{eqn:linear}) (purple) and eqn.~(\ref{eqn:exp}) (green).
    The cyan line shows 50 nm which corresponds to the size of the condensin.
    }
    \label{fig:x_with_time_direct}
 \end{center}
\end{figure}
\begin{figure}[H]
  \begin{center}
   \includegraphics[width=12cm]{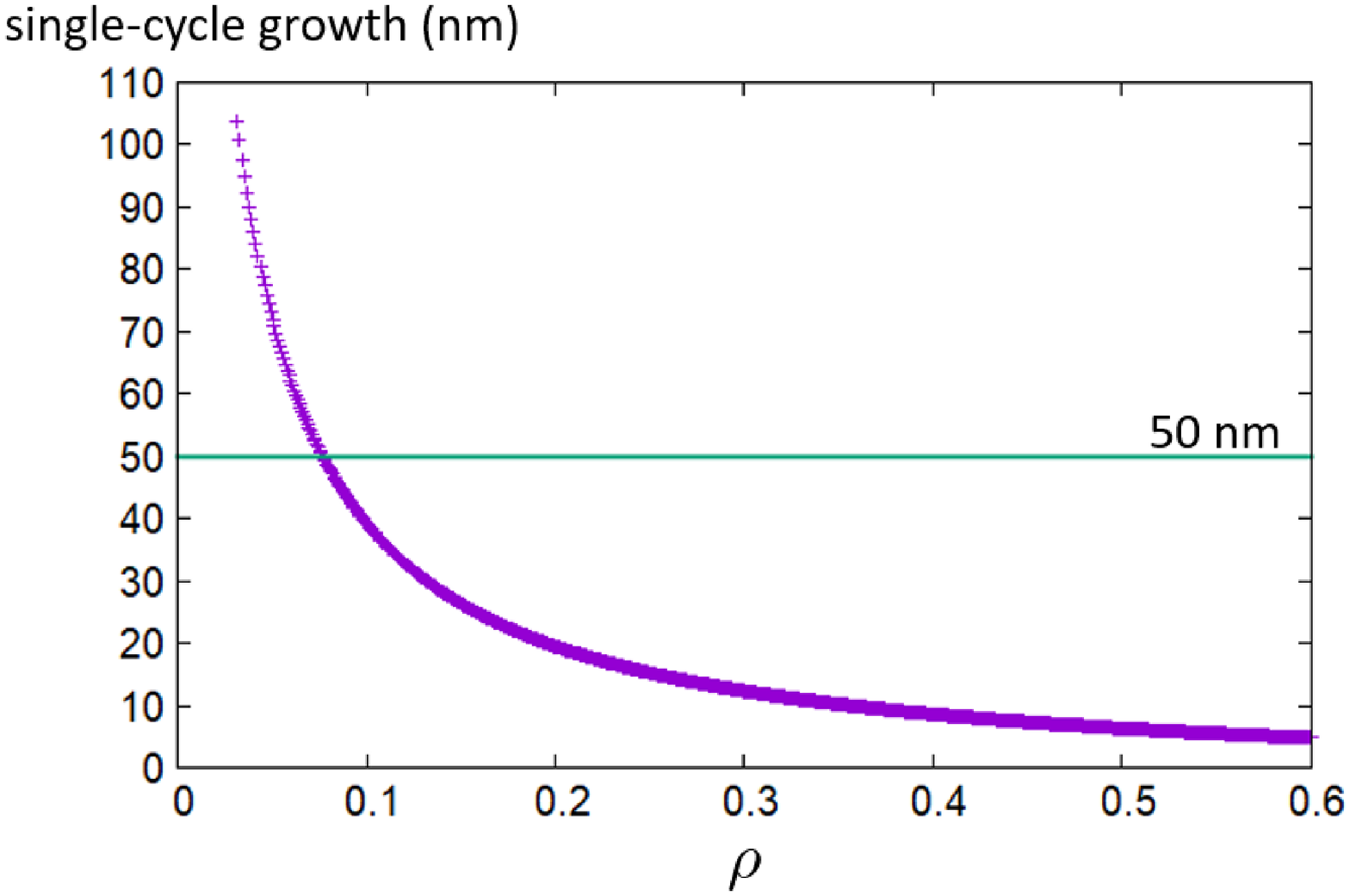}
    \caption{
    The purple line shows the dependence of the single-cycle growth on $\rho$ based on eqn.~(\ref{eqn:free_energy_direct}).
    The vertical and horizontal axes represent $x_{\rm int}$ and $\rho$, respectively.
    The green line shows 50 nm which is the size of the condensin.
    }
    \label{fig:x_with_time_direct}
  \end{center}
\end{figure}

When the nucleosome density $\rho \simeq 0.080 [/ b^3] =0.080 \times 10^{-3} [/ {\rm nm}^3]$ is achieved, the thermal-driving scenario is connected to the motor-pulling scenario.
As this value of the density is near 0, the perturbation expansion up to the lower order (the second order) might be justified.
Therefore, the crossover between the thermal-driving scenario and the motor-pulling scenario can be discussed using the perturbation theory, which we used in the main text.

\section*{Acknowledgments}
The authors thank T. Hatsuda and Y. Sakai for useful comments and discussions.
This work was supported by a Grant-in-Aid for Scientific Research (grant number 18H05276) from the Ministry of Education, Culture, Sports, Science, and Technology (MEXT), Japan. 
We partially performed the numerical computations using RIKEN\rq{} s supercomputer system HOKUSAI GreatWave.

\bibliography{../ref}
\bibliographystyle{aip}%
\end{document}